\documentclass{article}
\usepackage{spconf,amsmath,graphicx}
\usepackage[nolist]{acronym}
\usepackage{hyperref}

\usepackage{enumitem}
\setlist{nosep, leftmargin=14pt}

\usepackage{mwe} % to get dummy images

\usepackage{xspace}
\makeatletter
\DeclareRobustCommand\onedot{\futurelet\@let@token\@onedot}
\def\@onedot{\ifx\@let@token.\else.\null\fi\xspace}
\def\etal{\emph{et al}\onedot}

\title{Improved HER2 Tumor Segmentation with Subtype Balancing using Deep Generative Networks}
\name{
    \begin{tabular}{c}
        Mathias Öttl$^1$ , Jana Mönius$^2$, Matthias Rübner$^3$, Carol~I. Geppert$^2$, Jingna Qiu$^4$,\\Frauke Wilm$^{1,4}$, Arndt Hartmann$^2$, Matthias~W. Beckmann$^3$, Peter~A. Fasching$^3$, Andreas Maier$^1$,\\Ramona Erber$^2$, Katharina Breininger$^4$
    \end{tabular}
}

\address{
     $^1$ Pattern Recognition Lab, Friedrich-Alexander-Universität Erlangen-Nürnberg (FAU), Germany\\ 
     $^2$ Institute of Pathology, University Hospital Erlangen, FAU, Germany\\ 
     $^3$ Department of Gynecology and Obstetrics, University Hospital Erlangen, FAU, Germany\\
     $^4$ Department Artificial Intelligence in Biomedical Engineering, FAU, Germany
}

\begin{document}
\ninept
\maketitle
\begin{abstract}
Tumor segmentation in histopathology images is often complicated by its composition of different histological subtypes and class imbalance. Oversampling subtypes with low prevalence features is not a satisfactory solution since it eventually leads to overfitting. We propose to create synthetic images with semantically-conditioned deep generative networks and to combine subtype-balanced synthetic images with the original dataset to achieve better segmentation performance. We show the suitability of \acp{gan} and especially diffusion models to create realistic images based on subtype-conditioning for the use case of HER2-stained histopathology. Additionally, we show the capability of diffusion models to conditionally inpaint HER2 tumor areas with modified subtypes. Combining the original dataset with the same amount of diffusion-generated images increased the tumor Dice score from 0.833 to 0.854 and almost halved the variance between the HER2 subtype recalls. These results create the basis for more reliable automatic HER2 analysis with lower performance variance between individual HER2 subtypes.
\end{abstract}
\acresetall
\section{Introduction}
\label{sec:intro}

The extraction of distinct features to differentiate individual subtypes from one composite class can be problematic for machine learning algorithms, leading to a weakened performance for the main task and an inconsistent performance among subtypes~\cite{bias_1, subgrouping}.
In histopathology, tumors can be subtyped by origin or reason for growth \cite{breast_cancer_subtypes_1}. In this work, we focused on different histological subtypes among \ac{her2}-stained breast cancer samples, defined according to scoring systems for \ac{her2}. Each tumor cell in \ac{her2}-stained tissue can be scored as 0, 1+, 2+, or 3+, leading to the \ac{her2} tumor class being a composition of these subtypes~\cite{breast_cancer_her2_1}. An aggregated \ac{her2} score is usually assigned to each tumor sample, based on the composition of \ac{her2}-scored tumor tissue~\cite{breast_cancer_her2_2}. Treatment choices are based on the aggregated \ac{her2} score; thus, correct results for the subtype composition are essential when automatic tumor segmentation is employed. In this work, we consider the first step in a \ac{her2} segmentation pipeline, which is the segmentation of tumor tissue against background tissue. Different \ac{her2} subtype characteristics combined with a different prevalence of these \ac{her2} subtypes can lead to an inconsistent tumor segmentation performance between the underlying subtypes.

During algorithm development, this inconsistent performance across individual subtypes can be approached with oversampling of underrepresented subtypes, which can, however, quickly lead to overfitting due to the limited amount of training data available~\cite{oversamp}. To avoid overfitting, the training set can be extended by synthetic images created by a generative model, specifically \acp{gan}~\cite{gan}, which have been successfully applied for a wide range of medical applications. Chen \etal{} reviewed 105 publications in this domain, and for microscopic pathology, a performance increase was reported for all but one work~\cite{generative_1}.

Inspired by these results, we propose to use generative models to generate subtype-balanced synthetic tumor image datasets. Similar to the work of Fajardo \etal{}~\cite{generative_2}, we employ \acp{gan} for image generation, but extend it by semantic conditioning, i.e. a generative model is tasked to create an output that matches a two-dimensional label mask. Recently, diffusion models have shown great potential in image synthesis~\cite{diff_base} and have previously outperformed \acp{gan}~\cite{generative_3}. Therefore, we introduce diffusion models for semantic image synthesis in histopathology aiming to tackle subtype imbalances within our dataset. Furthermore, we experiment with partial image synthesis, where we use semantically-conditioned diffusion models to inpaint tumor regions, while the background remains unchanged.  

We investigate how different amounts of synthetic images from these three generative methods (\ac{gan}-generated, diffusion model-generated, diffusion model-inpainted) affect the subsequent tumor segmentation performance and how well the individual subtypes are segmented.\\
The main contributions of this paper include the following:
\begin{itemize}
  %\item Evaluation of how segmentation performance of tumor tissue in \ac{her2}-stained images is influenced by its subtypes.
  \item Illustration of the suitability of \acp{gan} and especially diffusion models to create realistic \ac{her2} images using semantic subtype-conditioning. Demonstration of the suitability of diffusion models for semantic subtype-conditioned inpainting in \ac{her2} histopathology images.
  \item Analysis of how different amounts of additional synthetic images influence the segmentation performance, specifically the overall tumor segmentation performance and the performance on individual subtypes.
\end{itemize}

\section{Dataset and Methods}
\label{sec:methods}

\subsection{Dataset} \label{sub_sec:dataset}

\begin{figure}[ht]
 \begin{minipage}[b]{1.0\linewidth}
  \includegraphics[width=\textwidth]{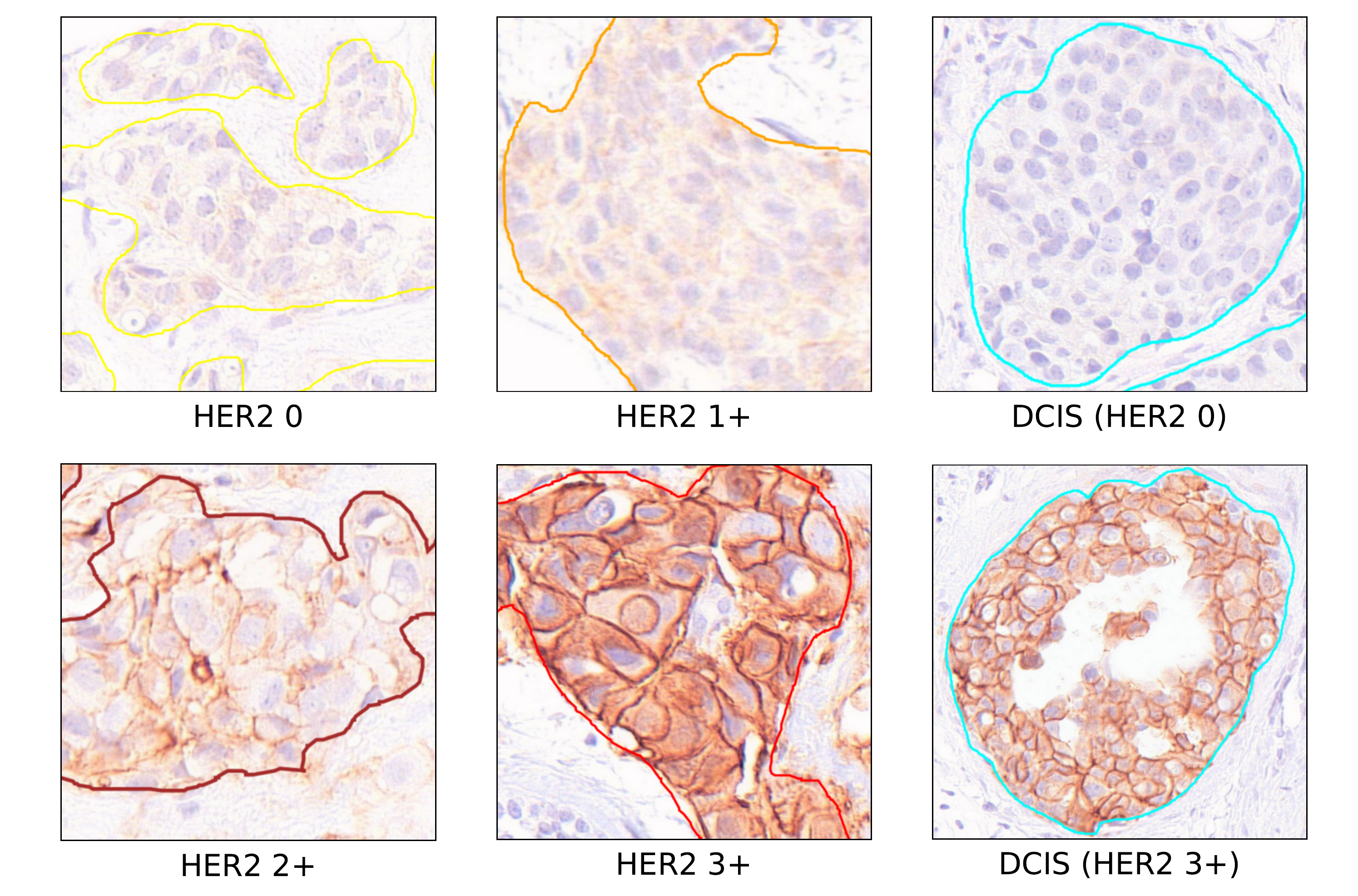}
 \end{minipage}
 \caption{The \ac{her2} tumor subtypes present in the \ac{her2} annotations, including two \ac{dcis} examples with tissue corresponding to different \ac{her2} subtypes.}
 \label{img:subType}
\end{figure}

\begin{figure}[ht]
 \begin{minipage}[b]{1.0\linewidth}
  \includegraphics[width=\textwidth]{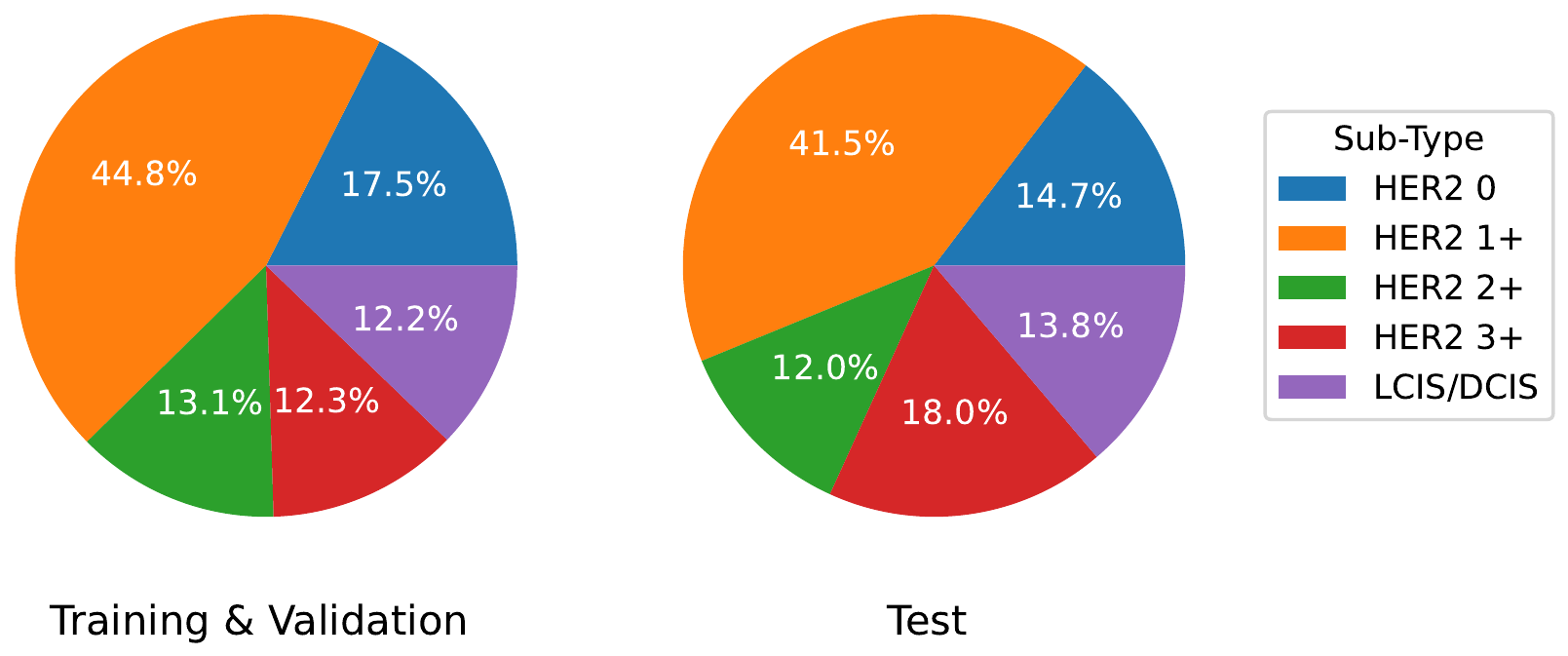}
 \end{minipage}
 \caption{Distribution of the different \ac{her2} subtypes across the annotated tumor regions for the combined training and validation set, as well as for the test set.}
 \label{img:subTypeDist}
\end{figure}

The data used in this work originated from 40 breast tissue sections from 40 different patients. The tissue was histochemically stained for \ac{her2} and digitized as \acp{wsi} with the 3DHistech PANNORAMIC 1000 scanner, using a $20\times$ objective lens. 
Despite methods to reduce the effort~\cite{bvm_sp}, annotation of \ac{wsi} is still not reasonable; thus, ten regions-of-interest of size 1.5\,mm$\times$1.5\,mm were selected from each \ac{wsi}. Cell groups of the same subtype were annotated as one tumor tissue instance using polygon contours on the EXACT platform~\cite{exact}. Five subtypes of tissue were considered: the four \ac{her2} subtypes according to the asco/cap guidelines~\cite{breast_cancer_her2_3} and, as a fifth subtype, \ac{lcis}/\ac{dcis} as one composite class. Figure~\ref{img:subType} shows examples of the annotated tissue types. \Ac{lcis} and \ac{dcis} describe non-invasive tumor tissues, which could be assigned to one of the four \ac{her2} classes. This assignment was not available; therefore, these classes were handled as a composition of the four \ac{her2} subtypes. Annotations were performed by a medical student and reviewed by a board-certified pathologist. A 24-8-8 train-validation-test split was carried out on \ac{wsi} level, with an equal amount of \ac{her2}-scored sections in each set. In Figure~\ref{img:subTypeDist}, the tumor tissue composition of the different dataset splits is shown. 

\subsection{Synthetic Image Generation}

\begin{figure*}[htb]
  \includegraphics[width=\textwidth]{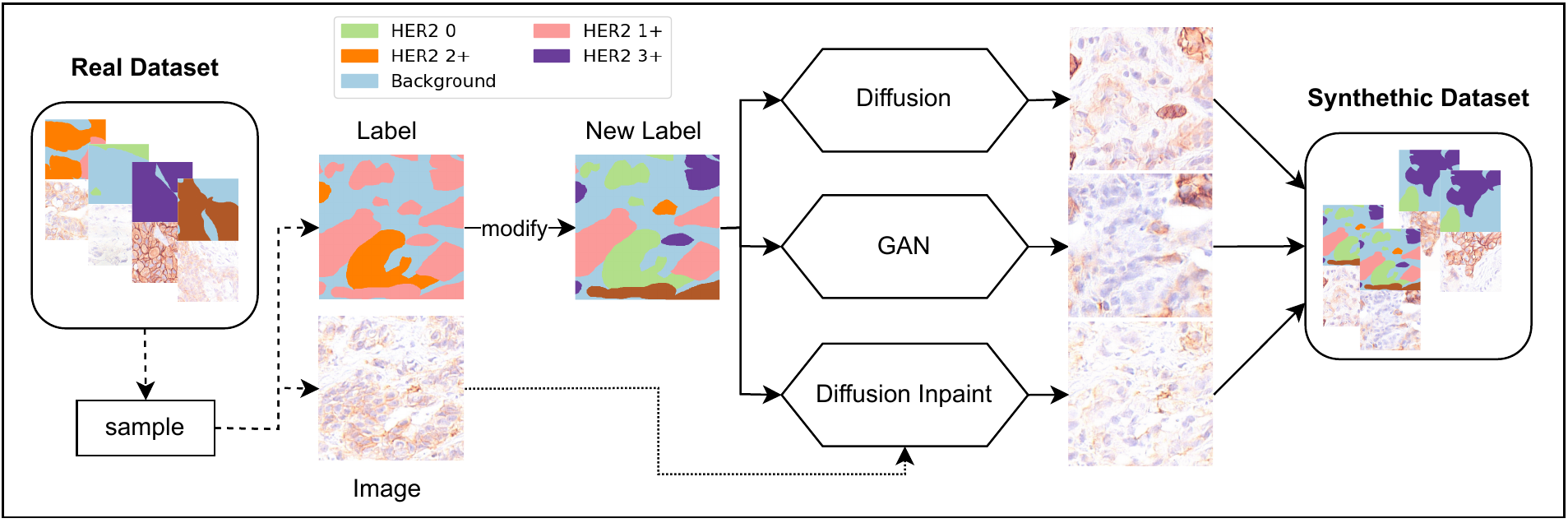}
  \caption{Illustration of how the synthetic datasets are created. An image and a corresponding label mask are sampled from the real dataset, and the subtype label of each tumor tissue instance is randomly modified. With the new label masks, synthetic images are created using a \ac{gan}, a diffusion model or diffusion inpainting. The generated images, together with their new label masks, are added to the synthetic dataset.}
  \label{img:overview}
\end{figure*}

Figure~\ref{img:overview} illustrates how synthetic images were created using semantic conditioning. As a first step, we modified the label masks, so that the resulting synthetic dataset was subtype-balanced. For this, we sampled an image patch and the corresponding label mask from our dataset and assigned a new, randomly selected subtype label to each tumor tissue instance. Due to a sufficiently high number of tumor tissue instances, the random assignment resulted in a subtype-balanced synthetic dataset. The modified label masks were then utilized for three methods of synthetic image generation as follows:

\textbf{\Ac{gan} image generation.} We used the \ac{gan}-based architecture proposed by Park \etal{} in \cite{spade} using spatially-adaptive normalization, which enables more realistic outputs for conditioning masks where only a single class is present. Sampling of such label masks is a common finding in our dataset; therefore, this technique ensures higher-quality synthetic images. To allow various outputs for the same conditioning, we used the latent space representation of a variational autoencoder as additional input, as proposed by Park \etal{} in the Appendix of their work. For image generation, a random vector was used as additional input.

\textbf{Diffusion model image generation.} We utilized latent diffusion models, as proposed by Rombach, Blattmann,~\etal{}~\cite{diff_1}. An autoencoder, which consists of a compressing encoder and decompressing decoder, is utilized to create a lower-dimensional latent representation of the data. The image generation process takes place in the latent space, which reduces computational cost due to the compression while achieving state-of-the-art performance. We considered this architecture advantageous as we expect it to better scale for generating large(r) image patches and large datasets, although we used relatively small patches of size $512\times512$ in this work. To create synthetic images, the latent diffusion model was utilized with the same modified label masks as the \ac{gan}.

\textbf{Diffusion model inpainting.} We used the latent diffusion model to inpaint the tumor in existing images, conditioned by the modified \ac{her2} subtypes. Unlike the above two methods, diffusion model inpainting only modifies the tumor tissue instances while keeping the background unchanged. Inpainting with diffusion models takes the context of an image into account, allowing the model to capture background characteristics when inpainting the new tumor tissue. Depending on how the background characteristics affect the data, this effect could be positive or negative for the subsequent task.

The generative models, as well as the subsequent segmentation task, were trained with the data described in Subsection~\ref{sub_sec:dataset}. We followed the \ac{gan} implementation provided by Park \etal{}~\footnote{\url{https://github.com/NVlabs/SPADE}}, which we trained with a batch size of 16 and the Adam optimizer (learning rate 1e-5). For latent diffusion, we adapted the implementation provided by Rombach, Blattmann, and colleagues~\footnote{\url{https://github.com/CompVis/latent-diffusion}}. The autoencoder for latent space computation was based on the provided vq-f4 configuration, which compressed the input by a factor of eight. The diffusion model was trained with a batch size of 16 and the Adam optimizer (learning rate 1e-6).

\subsection{Segmentation Network}

The segmentation aimed to analyze the effects of various quantities of synthetic training images from different generative methods. We used the U-Net architecture, proposed by Ronneberger \etal{} \cite{unet}, for segmentation,  which has proven itself a suitable architecture for segmentation in histopathology. For our experiments, we used a U-Net architecture with a ResNet-34~\cite{resnet} backbone that was pre-trained on ImageNet~\cite{imagenet}. The segmentation task consisted of two output classes (background, tumor) and was optimized with a combined Dice~\cite{dice} and cross-entropy loss. 
Hyperparameters were optimized on the tumor subtype-sampled dataset and retained for all experiments. The Adam optimizer (learning rate 1e-6) was utilized, with a batch size of 16.

\section{Experiments and Results}
\label{sec:results}

We evaluate the proposed methods both qualitatively in terms of a visual assessment of the generated images and quantitatively with regards to the resulting segmentation performance. For quantitative evaluation of the tumor segmentation, two metrics were considered. The tumor Dice score evaluates the segmentation result independent of subtype. The variance between the \ac{her2} tumor subtype recalls evaluates how largely the segmentation performance varies across the individual \ac{her2} subtypes. We will call this metric \textit{subtype variance} and lower values are favorable.

Different combinations of training data for the segmentation network were evaluated. The two baselines sampled all tumor tissue once independently of the subtypes (tumor sampled) and once uniformly across all tumor subtypes (subtype sampled). Different amounts of synthetic images extended the subtype-sampled dataset to measure the impact of synthetic data. In our experiments, we added 50\%, 100\%, 200\%, and 400\% of the original dataset size as additional synthetic images. All experiments were repeated five times to create reliable results.

%We first show qualitative results for the three methods of synthetic image generation with semantic conditioning. Then we introduce the different combinations of training data used to create the segmentation results, as well as the metrics used for analyzing the segmentation results. Finally, we present the improvement of the subtype segmentation performance for the best data combination.

\begin{figure}[ht]
 \begin{minipage}[b]{1.0\linewidth}
  \includegraphics[width=\textwidth]{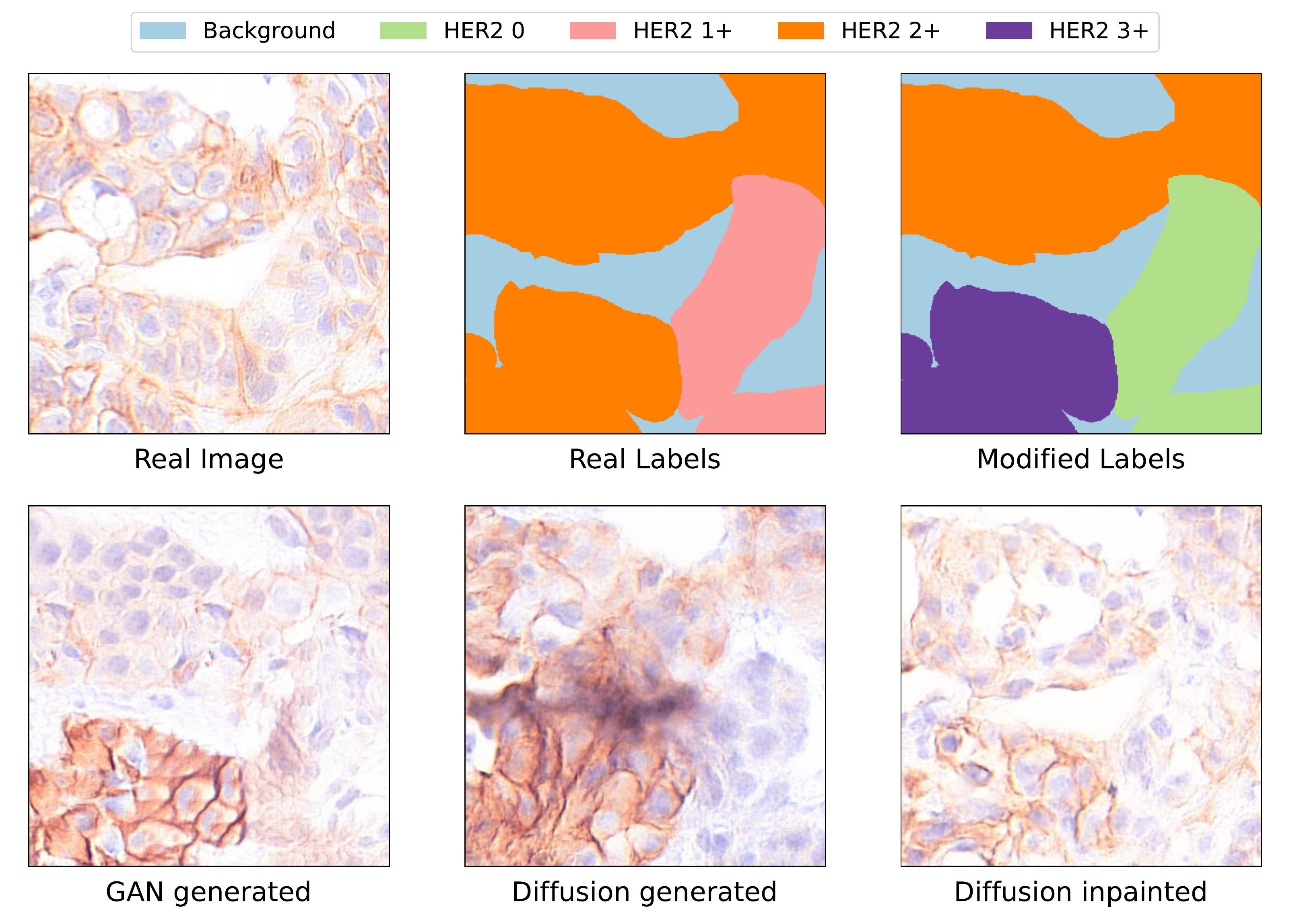}
 \end{minipage}
 \caption{Visual comparison of images created by generative networks.}
 \label{img:visExample}
\end{figure}

\begin{figure}[ht]
 \begin{minipage}[b]{1.0\linewidth}
  \includegraphics[width=\textwidth]{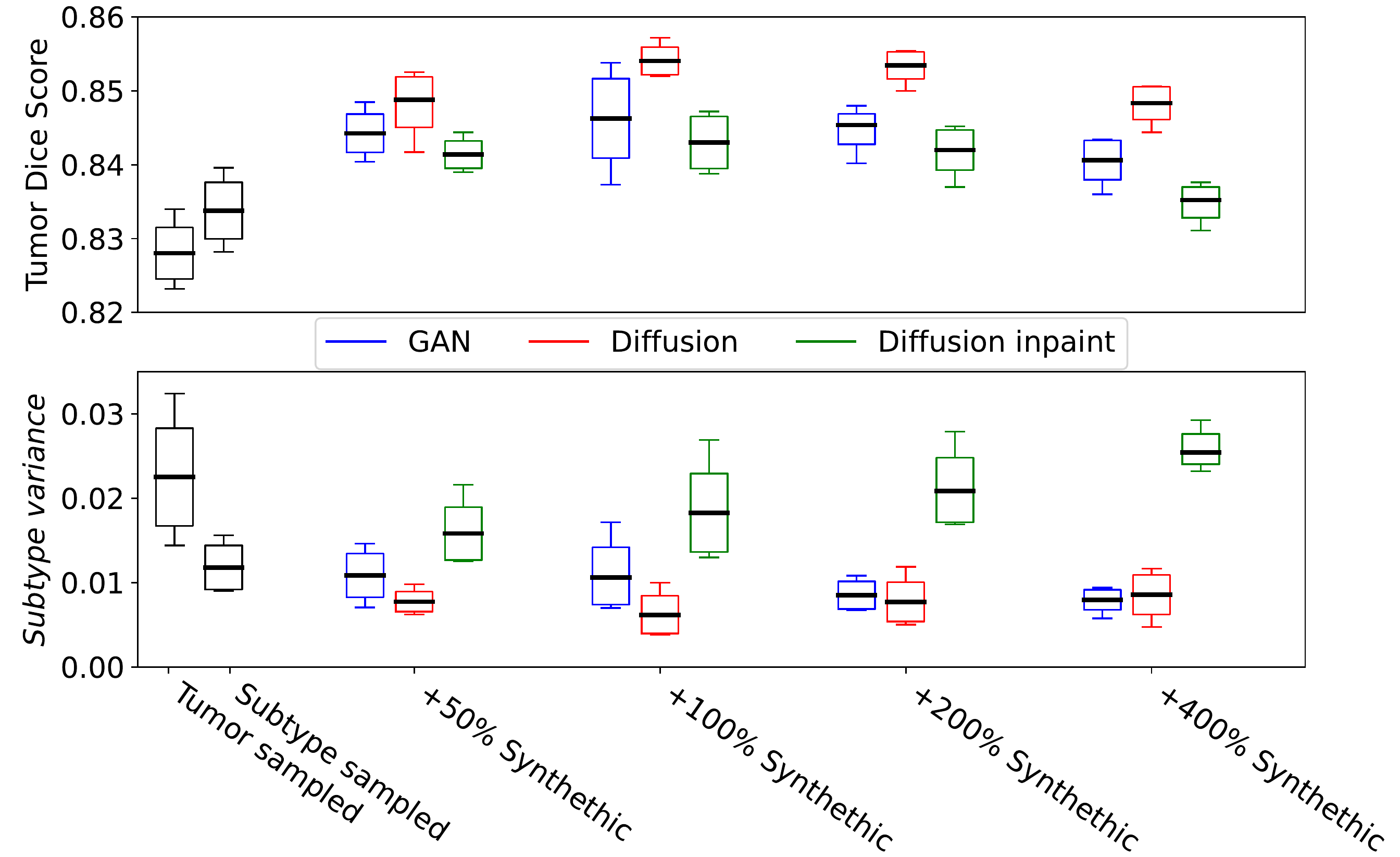}
 \end{minipage}
 \caption{Boxplots of the tumor Dice score and the \textit{subtype variance} for different configurations. Mean and standard deviation are visualized with the boxplot, while the whiskers mark the minimum and maximum values.}
 \label{img:metrics}
\end{figure}

\begin{figure}[htb]
 \begin{minipage}[b]{1.0\linewidth}
  \includegraphics[width=\textwidth]{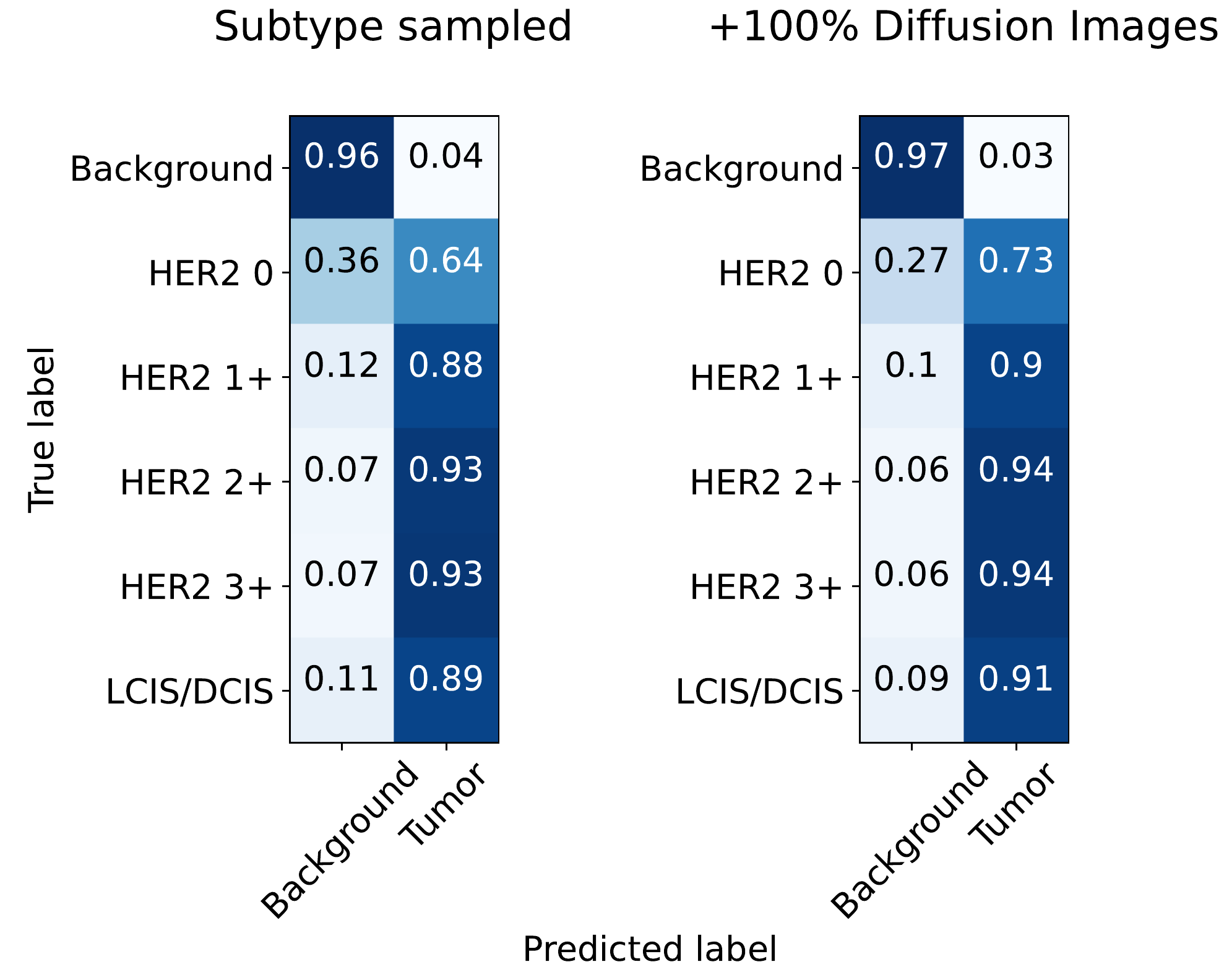}
 \end{minipage}
 \caption{Row normalized confusion matrix with the tumor subtypes. Left are the averaged values from the subtype-sampled runs, while right are the results from the experiment where 100\% diffusion images were added.}
 \label{img:conf}
\end{figure}

\subsection{Qualitative Results - Image Generation}

Figure~\ref{img:visExample} shows an example of synthetic images created with semantic conditioning. All three synthetic images are visually similar to real \ac{her2} histopathology images. The generated tumor structures show the staining characteristics of the subtype they were conditioned on. Signs of repeating patterns are visible in the \ac{gan}-generated images, and the background was often created as plain white without background structures. Diffusion-generated images show a high variation within all tissue types, and even rare background artifacts, like those seen in the example image, were generated. Diffusion-inpainted images also show a high variation within the created \ac{her2} subtypes, but the network favored the creation of tissue with staining closer to the original image.

\subsection{Quantitative Results - Tumor segmentation}

The performance metrics for the tumor segmentation are visualized in Figure~\ref{img:metrics}. Subtype sampling is superior to tumor sampling in both metrics. Adding synthetic data to the subtype-sampled data increased the tumor Dice score for all synthetic image methods, with diffusion-generated images performing best, \ac{gan}-generated images second, and diffusion-inpainted images last. 
For the \textit{subtype variance}, diffusion-inpainted images performed inferior to subtype sampling, while \ac{gan}-generated images and diffusion-generated images were superior. Diffusion-generated images achieved a better \textit{subtype variance} than \ac{gan}-generated images in all but one case. The best metrics were achieved when adding 100\% synthetic images, where diffusion-generated images increased the Dice score from 0.833 to 0.854 and decreased the \textit{subtype variance} by 47.8\%.

For the subtype-sampled dataset and the best-performing experiment, the averaged confusion matrices per subtype are shown in Figure~\ref{img:conf}. The most significant performance increase could be observed for the \ac{her2} 0 subtype, where the recall increased from 0.64 to 0.73. For the rest of the subtypes, minor improvements were achieved.

\section{Discussion}
\label{sec:discussion}

Although the qualitative results of the synthetically generated images look promising, some aspects remain open for discussion. \Ac{gan}-generated images show signs of repeating patterns, a common finding among \acp{gan}, which was not fully avoided with spatially-adaptive normalization~\cite{spade}. Additionally, the \ac{gan} network favored creating plain white background, which we suspect to be the network collapsing to the ``easiest'' solution for the background present in the training data. The diffusion-generated images were visually more compelling and even showed rare background structures, e.g. staining artifacts, which can help the segmentation network to become more robust toward these structures. 
%Realistic background structures are important since they help the network to learn stronger features for discriminating tumor tissue from background. In the given example, the network can learn that brownish staining is not always associated with tumor tissue. 
Diffusion model inpainting created tumor tissue with staining intensities more similar to the original image. We suspect that the remaining background information influenced the created staining levels in an unfavourable manner.

All presented methods for synthetic image generation improved the composite tumor segmentation, although no additional annotated images were available for the generative networks. We assume that generative networks might be able to interpolate between tissue features; for example, they could combine subtype invariant features, like tumor or cell shape, with subtype-specific features, like staining intensity. This could lead to synthetic tumor tissue, which has a combination of features that is not present in the original dataset. We suspect this to be the main reason for the lower \textit{subtype variance} when \acp{gan} and diffusion models are utilized.

The main performance benefit for diffusion models accrued for the \ac{her2} 0 subtype, which had a significantly lower recall than the other subtypes. We suspect that the lower recall was caused by the absence of the brownish staining for this class, which appears to be an easy indicator of tumor tissue. This indicates that the subtyping into four \ac{her2} classes might not be optimal and a subtyping into non-stained (\ac{her2} 0) and stained (\ac{her2} 1+, \ac{her2} 2+, \ac{her2} 3+) would be an interesting alternative. Diffusion models appeared to be able to correctly generate non-stained \ac{her2} tumor tissue for the learned representations, thus improving the performance for this subtype. Although we only consider tumor segmentation in this work, this more uniform performance between subtypes could lead to more reliable automatic \ac{her2} scoring in the future, since the scoring is based on the proportion of the \ac{her2} subtype tissue~\cite{breast_cancer_her2_2}.

These results are promising, but some limitations have to be noted. The same persons annotated training and test data, which could introduce a bias that affected evaluation metrics. Training of \acp{gan} is notoriously unstable and hard to monitor; therefore, it is possible that the model used in this work was not perfectly adapted to the data and could have produced better results.

\section{Conclusion}
\label{sec:conclusion}

We proposed to subtype balance \ac{her2} data with generative models. We showed the suitability of generative models, especially diffusion models, to generate semantically-conditioned synthetic images with a realistic appearance. Combining an equal amount of real images with diffusion model-generated images increased the Dice score of the tumor segmentation by 2.43\% and reduced the variance between the tumor subtype recalls by 47.8\%. This method is superior to oversampling subtypes and does not require additional annotated data.

The current approach requires the annotation of individual \ac{her2} to improve the tumor segmentation, and other kinds of so far non-annotated subtypes might exist. Future work could explore methods to alleviate this by, for example, unsupervised tumor area clustering and balancing these clusters.

Another future work could explore the use of fully synthetic training data. Besides the subsequent algorithm performance, one interesting aspect could be whether the synthetic images can be traced back to real patients. Such work could lay the foundation for using fully synthetic datasets and thereby reduce data privacy concerns.

% To start a new column (but not a new page) and help balance the last-page
% column length use \vfill\pagebreak.
% -------------------------------------------------------------------------

\section{Compliance with ethical standards}
All procedures performed in studies involving human participants were in accordance with the ethical standards of the institutional and/or national research committee and with the 1964 Helsinki declaration and its later amendments or comparable ethical standards.

\section{Acknowledgement}
This project is supported by the Bavarian State Ministry of Health and Care, project grants No. PBN-MGP-2010-0004-DigiOnko and PBN-MGP-2008-0003-DigiOnko. We also gratefully acknowledge the support from the Interdisciplinary Center for Clinical Research (IZKF, Clinician Scientist Program) of the Medical Faculty FAU Erlangen-Nürnberg. K.B. and J.Q. gratefully acknowledge support by Dhip campus - Bavarian aim.

% References should be produced using the bibtex program from suitable
% BiBTeX files (here: strings, refs, manuals). The IEEEbib.bst bibliography
% style file from IEEE produces unsorted bibliography list.
% ------------------------------------------------------------------------- 
\bibliographystyle{IEEEbib}
\bibliography{strings,refs}

\begin{acronym}[type=hidden] 
\acro{her2}[HER2]{Human Epidermal growth factor Receptor 2} 
\acro{gan}[GAN]{Generative Adversarial Network}
\acro{wsi}[WSI]{Whole Slide Image}
\acro{dcis}[DCIS]{Ductal Carcinoma In Situ}
\acro{lcis}[LCIS]{Lobular Carcinoma In Situ}
\acro{tpr}[TPR]{True Positiv Rate}
\end{acronym}

\end{document}